\documentclass[%
 aip,
 amsmath,amssymb,
 reprint,%
]{revtex4-1}

\usepackage{graphicx}
\usepackage{dcolumn}
\usepackage{bm}

\usepackage[utf8]{inputenc}
\usepackage[T1]{fontenc}
\usepackage{mathptmx}
\usepackage{etoolbox}
\usepackage{natbib}

\makeatletter
\def\@email#1#2{%
 \endgroup
 \patchcmd{\titleblock@produce}
  {\frontmatter@RRAPformat}
  {\frontmatter@RRAPformat{\produce@RRAP{*#1\href{mailto:#2}{#2}}}\frontmatter@RRAPformat}
  {}{}
}%
\makeatother
\begin{document}

\preprint{AIP/123-QED}

\title[Advancing wave fronts retarded by a rod canopy after a dam break]{The advancing wave front on a sloping channel covered by a rod canopy following an instantaneous dam break}

\author{Elia Buono}%
\email{elia.buono@polito.it}%
\affiliation{Dipartimento di Ingegneria dell'Ambiente, del Territorio e delle Infrastrutture, Politecnico di Torino, Torino, Italia}%
\author{Gabriel G. Katul}%
\email{gaby@duke.edu}%
\affiliation{
 Department of Civil and Environmental Engineering, Duke University, Durham, North Carolina, USA}%
\author{Davide Poggi}%
 \email{davide.poggi@polito.it}%
\affiliation{Dipartimento di Ingegneria dell'Ambiente, del Territorio e delle Infrastrutture, Politecnico di Torino, Torino, Italia}%

\date{\today}


\begin{abstract}
The drag coefficient $C_d$ for a rigid and uniformly distributed rod canopy covering a sloping channel following the instantaneous collapse of a dam was examined using flume experiments. The measurements included space $x$ and time $t$ high resolution images of the water surface $h(x,t)$ for multiple channel bed slopes $S_o$ and water depths behind the dam $H_o$ along with drag estimates provided by sequential load cells.  Analysis of the Saint-Venant Equation (SVE) for the front speed using the diffusive wave approximation lead to a front velocity $U_f=\sqrt{\Gamma_h 2 g \phi_v'/(C_d m D)}$, where $\Gamma_h=-\partial h/\partial x$, $g$ is the gravitational acceleration, $\phi_v'=1-\phi_v$ is fluid volume fraction per ground area, $\phi_v=m \pi D^2/4$ is the solid volume fraction per ground area, $m$ is the number of rods per ground area, and $D$ is the rod diameter.  An inferred $C_d=0.4$ from the $h(x,t)$ data near the advancing front region,  also confirmed by load cell measurements, is much reduced relative to its independently measured steady-uniform flow case.   This finding suggests that drag reduction mechanisms associated with transients and flow disturbances are more likely to play a dominant role when compared to conventional sheltering or blocking effects on $C_d$ examined in uniform flow.  The increased air volume entrained into the advancing wave front region as determined from an inflow-outflow volume balance partly explains the $C_d$ reduction from unity. 

\end{abstract}

\maketitle

\section{Introduction}
\label{Intro}
The sudden release of water following an instantaneous collapse of a dam has received much research attention in hydrology (e.g. overland flow), ecology (e.g. rapid inflow into wetlands or a marsh), hydraulics (e.g. flood routing), and coastal engineering (tsunami on coastal plains) for well over a century  \citep{ritter1892fortpflanzung,whitham1955effects,hunt1982asymptotic,hogg2004effects,daly2004similarity,chanson2006tsunami,begnudelli2007simulation,Ajayi2008,carrivick2010dam,chanson2009application,thompson2011unsteady,ma2012real}.  The hydrodynamics describing the unsteady and shallow nature of such flows are summarized by the Saint-Venant equation (SVE) introduced in 1871 \citep{de1871th,hager2019correspondence}.  For a rectangular prismatic section of constant width $B$, the SVE in their one-dimensional form are given by two partial differential equations: the continuity and the area-averaged momentum balance.  For the dam break problem, the SVE is expressed as  \citep{lighthill1955kinematic,whitham1955effects,french1985open,hogg2004effects,bellos19871}
\begin{equation}
\label{Cont}
\frac{\partial h}{\partial t}+\frac{\partial U h}{\partial x}=0,
\end{equation} 
and
\begin{equation}\label{SVE}
\frac{\partial U}{\partial t}+U \frac{\partial U}{\partial x}+g \left(\frac{\partial h}{\partial x}+S_f-S_o\right)=0,
\end{equation} 
where $x$ is the longitudinal distance from the dam location with $x=0$ set at the dam location, $t$ is time with $t=0$ set to the instant the dam is removed, $h(x,t)$ is the water depth, $U(x,t)$ is the area-averaged or bulk velocity, $g$ is the gravitational acceleration, $S_o$ is the bed slope, and $S_f$ is the friction slope that is unknown in the SVE.  It is the closure model for $S_f$ in the SVE that frames the scope of the work.  As early as 1892, analytical results for the dam-break problem were derived when $S_f=0$ and $S_o=0$ \citep{ritter1892fortpflanzung,ancey2008exact}.  The inclusion of finite $S_o$ but keeping $S_f=0$ revises the classical Ritter solution to 
\citep{chanson2009application,larocque2012experimental,melis2019resistance}
 \begin{equation}
\label{Ritter_U}
U(x,t)=\frac{2}{3}\left(\frac{x}{t}+U_o +S_o g t\right),
\end{equation} 
and
\begin{equation}
\label{Ritter_h}
h(x,t)=\frac{1}{9g}\left(2 U_o-\frac{x}{t}+\frac{1}{2}S_o g t\right)^2,
\end{equation} 
where $U_o=\sqrt{g H_o}$ is the initial celerity speed.  Here, the initial conditions to the SVE are a dry channel bed.  When $S_o=0$ and $t>0$, equations \ref{Ritter_U} and \ref{Ritter_h} can be expressed in a dimensionless and compact form as \citep{ritter1892fortpflanzung} 
\begin{equation}
\label{Ritter_n}
h_n=\frac{1}{9}\left(2-u_n\right)^2,
\end{equation} 
where $h_n=h/H_o$ is the dimensionless water depth, $u_n=(x/t){(U_o)}^{-1}$ is the dimensionless wave speed, $t_n=t (H_o/g)^{-1/2}$ is dimensionless time, and $x_n=x/H_o$ is dimensionless longitudinal position downstream from the dam.  Revisions to these results are numerous and include a gradual breaching of the dam \citep{capart2013analytical}, lateral contractions \citep{kocaman2012effect,kocaman2020experimentalCG,cozzolino2017exact}, asymmetric geometry \cite{ferdowsi2021development,wang2019comparison, wang2015analytical,wang2020analytical}, steep $S_o$ \citep{ancey2008exact,wang2015analysis}, and introduction of bends along the channel \citep{frazao2002dam}.  Perhaps the most studied revision to the inviscid solution is finite wall-friction  \citep{dressler1952hydraulic,whitham1955effects,hunt1982asymptotic,hunt1984perturbation,hunt1984dam,chanson2009application,wang2015analysis,nielsen2022friction}.  In prior applications with finite wall friction, a resistance formulation must be introduced to link $S_f$ to the variables being modeled by the SVE ($h$ or $U$).  The primary restriction imposed on such models are recovering outcomes based on locally steady and uniform flow conditions.  For these idealized flow conditions, the most common formulation to parameterize $S_f$ is Manning's formula \citep{Manning} that assumes a constant roughness coefficient (=$n$) and links $S_f$ to the sought variables using  
\begin{equation}
\label{Manning_Sf}
S_f=\left[\frac{2 g n^2}{R_h^{4/3}}\right] \frac{{U}^2}{2 g},
\end{equation} 
where $R_h=A_c/P_w$ is the hydraulic radius, $A_c=B h$ is the cross-sectional area of the flow, $P_w=B+2h$ is the wetted perimeter, and $n$ is Manning's roughness coefficient (in s m$^{-1/3}$ when SI units are used).  This formulation is common given the accrued information on $n$ over the years for different surface cover types \citep{Chow1959, french1985open}.  Also, theoretical justification for equation \ref{Manning_Sf} using turbulence theories and the energy cascade are also emerging \citep{Gioia2001,bonetti2017manning}.  In some applications, the kinematic wave approximation is invoked whereby the momentum balance is reduced to $S_f=S_o$ \citep{woolhiser1967unsteady}.  Invoking this approximation and inserting equation \ref{Manning_Sf} into the continuity equation, a broad class of self-similar solutions can be derived and connections between the dam-break equations and the Fokker-Planck equations have already been proposed \citep{daly2004similarity}. However, extending such solutions to the general SVE remains fraught with difficulties. 

Controlled laboratory experiments on this topic remain also limited despite the undisputed societal significance of the dam break problem \citep{chanson2009application,larocque2012experimental,ozmen2011dam}.  Some laboratory studies considered (i) single isolated obstacles \citep{soares2008dam,del2022experimental,di2020interaction,kamra2019experimental,ansari2021experimental,vosoughi2022downstream,kocaman2020experimentalIO,huo2023experimental,tan2023experimental}, (ii) the initial stages of an instantaneous dam-break over smooth surfaces \citep{stansby1998initial,ozmen2010dam,espartel2021experiments}, (iii) the use of polymer additives for inducing reductions in $S_f$ \citep{janosi2004turbulent}, (iv) geometric alterations to the channel section such as contractions, expansions, embankments and bends \citep{frazao2002dam,kocaman2012effect, kocaman2020experimentalCG,di2018dam}, and (v) the role of sediments and movable beds on $S_f$ \citep{abderrezzak2008one,zech2008dam,vosoughi2020experimental,biswal2018effects,khosravi2021laboratory,fent2019dam,qian2018new}.  However, the dam-break problem for channels covered by vegetation remains under-studied with less than a handful being reported \citep{melis2019resistance}.  When the channel is vegetated, explicit inclusion of distributed drag into the SVE is necessary as energy losses are no longer related to wall friction  \citep{nepf1999drag,wu1999variation,lawrence2000hydraulic,green2005modelling,huthoff2007analytical,poggi2009hydraulic,huai2009three,kothyari2009drag,nepf2012flow,etminan2017new}.  

The work here explores experimentally the effects of canopy drag on the physics of the advancing front following the instantaneous removal of a dam for varying static water depth $H_o$ behind the dam and $S_o$.  The canopy used is composed of staggered rigid cylinders covering the flume base downstream from a dam where $S_o$ is varied from $S_o=0$ to 3$\%$.  Attention is drawn to the role of canopy drag reduction mechanisms as the advancing front traverses the rod canopy. Thus, the two experimentally controlled variables to be manipulated here are $S_o$ and $H_o$.  Comparison with a prior study \citep{melis2019resistance} where the rod density was much higher is also presented. 

\begin{figure*}[ht!]
\noindent\includegraphics[width=40pc]{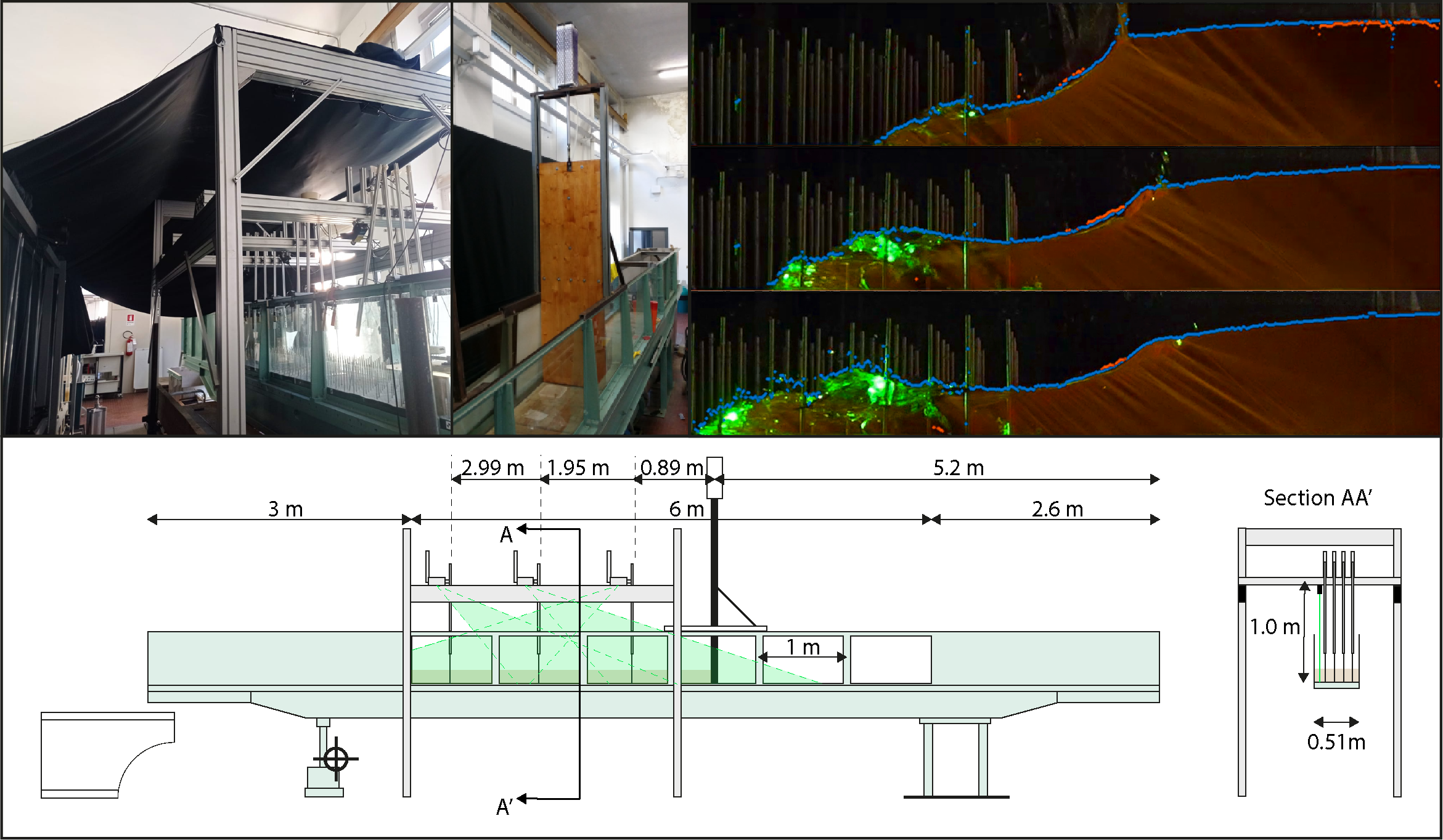}

\caption{The experimental setup for the dam break problem. On the top left, a picture of the channel downstream of the cofferdam and a picture of the wooden cofferdam are featured. On the top right, three images from the early stages of the dam break wave propagation ($H_o=$ 0.15m, $S_o=$ 0$\%$) are shown along with the detected free surface (blue dots). On the bottom, side view of the whole experimental facility (on the left) and longitudinal section view (on the right) with the most meaningful dimensions}
\label{Fig:setup}
\end{figure*}

\section{Theory}
\label{Theory}
The setup considered here is for an instantaneous removal of a dam that results in a flood wave propagating downstream along a sloping rectangular channel.  The channel is covered by a uniform rigid cylindrical rod canopy that acts to remove energy and momentum from the advancing flood wave.   The cylinders are staggered and presumed to have a uniform diameter $D$ and height $h_c/h(x,t)>1$ after the dam break.  The goal is to describe the front position $x_f$ and front speed $U_f$ downstream from the dam for various combinations of control variables $S_o$ and $H_o$. To arrive at an expression for $S_f$ that accounts for the presence of cylinders to be used in the SVE, a starting point is to consider a \textit{locally} steady-uniform flow within a canopy.  The canopy is presumed to be sufficiently dense so that ground and sidewall friction contributions to the total stress can be ignored.  Thus, a \textit{local} balance between the gravitational contribution of the water weight along $x$ and the drag resisting this motion results in  
\begin{equation}
\label{Force_Balance}
\rho g S_f V_w=C_d A_v \rho \frac{U^2}{2 g},
\end{equation} 
where $\rho$ is the water density, $V_w$ is the water volume, $A_v$ is the frontal area of the vegetation contained in $V_w$ and $C_d$ is the drag coefficient.  For the SVE, a force balance per unit ground area is preferred so that $V_w=h (1-\phi_v)$ and $A_v=m D h$, where $\phi_v$ is the solid volume fraction per ground area determined by $\phi_v=m \pi D^2/4$, $m$ is the rod density (i.e. number of rods per unit ground area). This force balance leads to 
\begin{equation}
\label{Drag_Sf}
S_f=\left[\frac{(C_d)~ m D}{1-\phi_v} \right] \frac{{U}^2}{2 g}.
\end{equation} 
Equations \ref{Manning_Sf} and \ref{Drag_Sf} can be made equivalent when introducing a non-constant Manning roughness given by
\begin{equation}
\label{Drag_n}
n=\sqrt{\frac{C_d m D}{2 g (1-\phi_v)}} R_h^{2/3}.
\end{equation} 
Setting $n$ to a constant value in models of $S_f$ cannot be reconciled with a distributed drag formulation.  The $C_d$, which frames the scope of the work here, is influenced by numerous interactions between the canopy elements and the moving water.  In steady-uniform flows, $C_d$ is presumed to vary with Reynolds number $Re=V L/\nu$, where $V$ and $L$ are characteristic velocity and length scales respectively, and $\nu$ is the water kinematic viscosity.  A number of possibilities have been introduced in the literature to define $L$ and $V$ in this context.  Some set $L$ to be proportional to $D$, rod spacing, or  $R_h$.  Likewise, $V$ was set to pore-scale velocity, the constricted velocity, or a separation velocity \citep{etminan2017new}.  Corrections such as sheltering or blockage due to the presence of an array of cylinders have also been studied for an isolated cylinder and an array of cylinders  \citep{tanino2008laboratory, stoesser2010turbulent,nepf1999drag, lee2004drag,etminan2017new,cheng2010hydraulic,Zdravkovich2000flow}. 

Returning to the water level description in $x$ and $t$ of an advancing wavefront within a rod canopy, a number of simplifications have been adopted to the SVE. Within the wavefront region, the front speed attains a quasi-constant value so that the unsteady and inertial terms $\partial U/\partial t + U \partial U/\partial x$ are small relative to remaining terms \citep{chanson2009application}.  For these standard simplifications, the SVE and the continuity equation become \citep{melis2019resistance}
\begin{equation}
\label{SVE_Diffusive_Wave}
g\left(\frac{\partial h}{\partial x}+S_f-S_o\right)=0; ~~ \frac{\partial h}{\partial t}+ U\frac{\partial h}{\partial x}=0.
\end{equation} 
At very high $Re$, $C_d$ may attain a quasi-constant value so that  
\begin{equation}
\label{SVE_Diffusive_Wave_Speed}
U=\sqrt{-A\left(\frac{\partial h}{\partial x}-S_o\right)},~ A=\frac{2 g(1-\phi_v)}{C_d m D}.
\end{equation} 
Inserting $U$ into the approximated continuity equation and solving for $h(x,t)$ results in 
\begin{equation}
\label{eq:h_wavetip_So}
h(x,t)=C_1+{C_2} t+ \frac{x}{3}\left[ S_o + E(S_o, A) \right],
\end{equation} 
where $E(S_o, A)$ is given by
\begin{align}
\label{h_wavetip_So2}
E &=\frac{R_1}{A} + \frac{A {S_o}^2}{R_1}; \rm{where}\\
R_1&=\left({A^3 {S_o}^3-\frac{27A^2 C_2^2}{2}+\frac{3\sqrt{3}}{2} 
   \sqrt{27 A^4 C_2^4-4 A^5 C_2^2 {S_o}^3}}\right)^{1/3}.\nonumber
\end{align} 
Here, $C_1$, and $C_2$ are integration constants independent of $x$ or $t$.  Equation \ref{h_wavetip_So2} applies when $S_o>0$. For $S_o=0$, the solution to the simplified continuity and SVE system is
\citep{melis2019resistance}
\begin{equation}
\label{eq:h_wavetip}
h(x,t)=C_1+{C_2} t- {\left[C_2 {\sqrt{\frac{C_d m D}{2g (1-\phi_v)}}} \right]}^{2/3} x.
\end{equation} 
It is to be noted that equation \ref{eq:h_wavetip} is not recovered from equation \ref{eq:h_wavetip_So} when setting $S_o=0$ as this condition resembles a singular limit (i.e. addition of a new force).  A near constant $C_d$ implies $h(x,t)$ is linear in $x$ (and $t$) at the advancing front region.  The slope of this linear dependence on $x$ varies with $C_d^{1/3}$ (i.e. sub-unity exponent). 

\begin{figure*}[ht!]
\noindent\includegraphics[width=40pc]{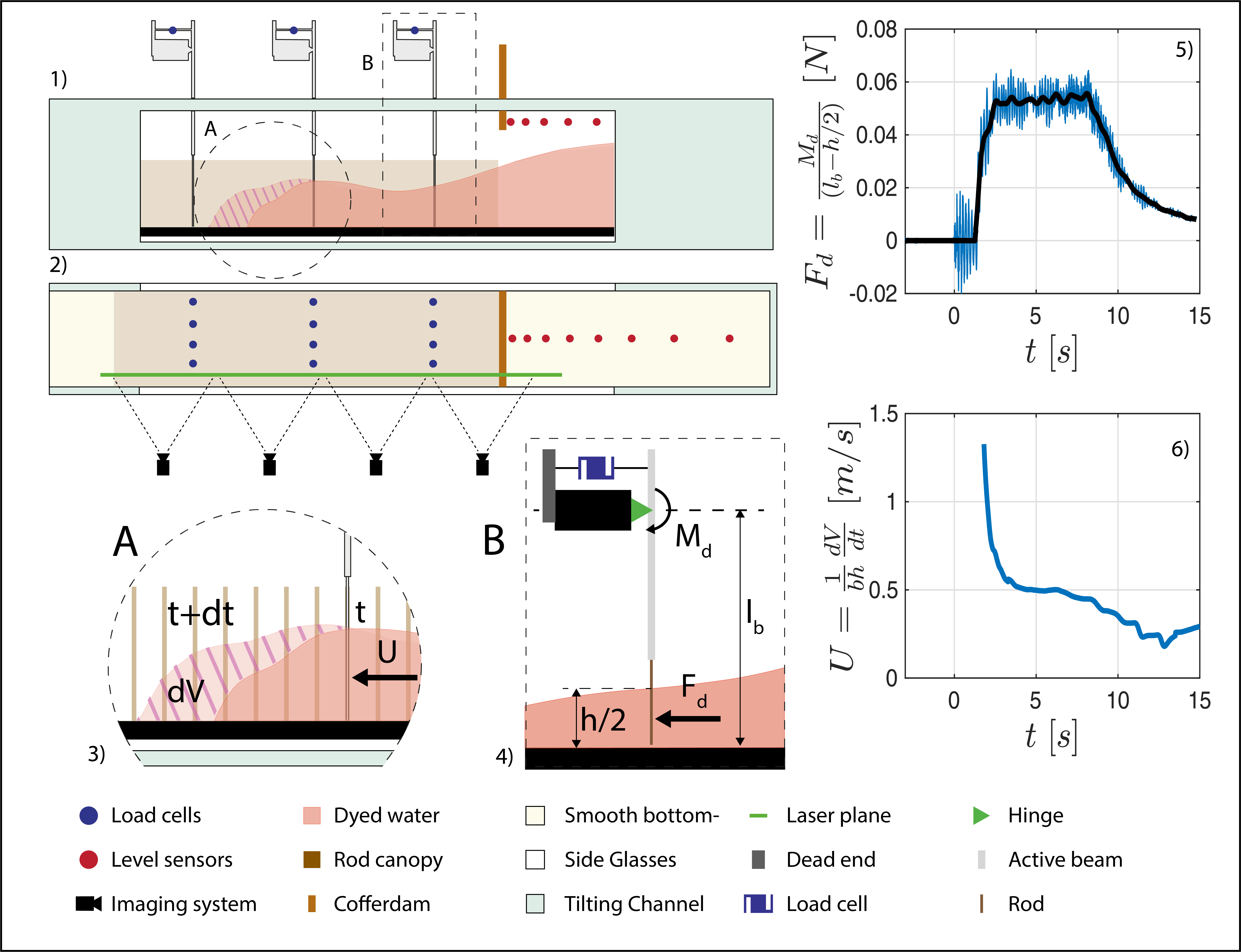}
\caption{A schematic of the instrument setup. 1) side view of the channel with the dam released, 2) top view of the channel indicating the positions of the measuring instrumentation, 3) detailed view of the advancing wave front showing how the bulk velocity from imaged water depth was computed from water level measurements at time $t$ and $t+dt$, 4) detailed view of the load cell system illustrating the method for drag measurements including the assumption of setting $F_d$ at $h/2$, 5) a sample drag time series record in raw (blue line) and filtered (black line) form for $H_o=$ 0.15m, $S_o=$ 0$\%$, $X_{lc}=$ 1.95m, $Y_{lc}=$ 0.05m, and
6) the computed bulk velocity (blue line) for $H_o=$ 0.15m, $S_o=$ 0$\%$, $X_{lc}=$ 1.95m.}
\label{Fig:setup2}
\end{figure*}

\section{Experiments}
\label{Experiments}
The flume facility has been described in prior studies \citep{melis2019resistance,poggi2004effect} and will not be fully repeated. Briefly, the tilting channel, the wooden cofferdam, the pneumatic pump release mechanism for the dam removal (mimicking an instantaneous dam break), the rod canopy, the water level imaging system, the load cells, the water level sensors, and the data acquisition system are featured in Figure \ref{Fig:setup}. The rectangular channel shown in this figure has a length $L=11.6$m, a width $B=0.51$m, and a side height $L_s=0.6$m.  The channel sides are made of glass to allow imaging and optical access.  The $S_o$ was varied from 0\% to 3\%.  The water behind the dam was filled until the target $H_o$ is reached ($H_o=$ 0.15m, 0.20m, 0.25m, 0.30m).  The quasi-instantaneous dam removal was carried out using a pneumatic pump that pulls rapidly the cofferdam vertically upward.  After the dam removal, the water discharges from the end of the channel into a recirculating tank while passing over a rectangular weir.  Downstream from the dam, an array of rigid cylinders of $D$=0.006 m and $h_c$=0.14 m were used to represent the vegetation. The cylinders were fixed onto boards attached to the channel bottom and cover an entire cross-section. A staggered rod configuration was used for all runs with a constant density $m=194$ rods m$^{-2}$.  

To image $h(x,t)$ during and after the dam removal, four synchronized Sony Handycam FDR-AX700 cameras were employed. The spatial resolution of each camera was 1920 $\times$ 1080 pixels interrogated in time at 100 frames per second.  The cameras cover a total length of 4.2 m starting from 0.6 m behind the cofferdam.  Water was mixed with a Rhodamine dye and a green laser plane was seated up parallel to the channel to enhance the automated detection of $h(x,t)$ and the delineation of the water surface profile.  MATLAB (Mathworks, Natick, Massachusetts, USA) was used to analyze the movies by transforming the detected $h(x,t)$ from pixel to metric coordinates as described elsewhere \citep{melis2019resistance}.  The duration of each experiment (i.e. a combination of $H_o$ and $S_o$) ranged from 5 to 10 s. 


Twelve load cells were used to record the drag force in $t$ at 1kHz on 3 downstream cross-sections away from the dam ($X_{lc}=$ 0.89m, 1.95m, 2.99m). On each section, four load cells were placed with 0.1m spacing starting at 0.05m from the left side, according to the staggered canopy's pattern. The load cells used were eight Leane model DBBSM-1kg-003-000 along with four Instrumentation Devices (model kD40s). The drag $F_d$ exerted on the instrumented rod is transferred to the load cell through a rigid active beam hinged on a fixed point as shown in Figure \ref{Fig:setup2}. The load cells were calibrated by applying a known torque to the cell-beam system. During the experiments, the torque onto the cell-beam system ($=M_d$) by $F_d$ was recorded. Then, the actual $F_d$ was derived assuming $F_d$ is concentrated at half the depth $F_d=M_d/(l_b-h/2)$ where $l_b$ is the distance from the channel bottom to the hinge. Additionally, eight level sensors (Balloff model BUS004W) were employed to record the water depth in $t$ at 1kHz behind the cofferdam.  These sensors were located at the cross-section center and spaced as $X_{ls}=$ -0.2m, -0.4m, -0.7m, -1m, -1.6m, -2m, -3m, -4m, where the free surface cannot be detected with the imaging system. An acquisition card (National Instruments, Austin, Texas, USA) was used for both load cells and level sensors. LabVIEW (National Instruments, Austin, Texas, USA) was used to drive data acquisition. 

The $C_d$ was quantified from the measured $F_d$ using the quadratic drag-law $F_d=(1/2) C_d (D h) U^2$.  This quantification requires the bulk velocity on the instrumented cross-section $U_{lc}$ that was not directly measured. For each instrumented cross section $X_{lc}$, the $U_{lc}$ was computed from measured $h(x,t)$ using the continuity equation. Specifically, for each $t$ the volume $V_{lc}$ forwarding $X_{lc}$ was computed by numerical integration of $h(x,t)$ over $x$.  The $V_{lc}$ was then numerically differentiated to obtain the flow rate $Q_{lc}$ through the instrumented section.  The last step is to determine the bulk velocity from $U_{lc}=Q_{lc}/(h(x=X_{lc},t) B)$. In Figure \ref{Fig:setup2}, an illustration of how the bulk velocity was computed from this procedure is provided. 

The dam break experiments reported here were compared to prior experiments conducted in the same flume, same $S_o$-$H_o$ combinations, and for the same staggered rod configuration  \citep{melis2019resistance}. The main difference between the present and the prior experiments is the rod density $m$.  In the prior experiments, $m=1206$ rods m$^{-2}$ whereas here $m=194$ rods $m^{-2}$.  These prior experiments did not include load cell measurements or independent water level measurements behind the dam. Hence, their effective $C_d$ was only inferred by fitting a numerical solution of the SVE to imaged $h(x,t)$ for all the $H_o$-$S_o$ combinations with assumptions about the inflow volume into the channel following the dam break.  The most pertinent finding from these prior experiments was that a $C_d=0.4$ better describes the measured $h(x,t)$ than the numerous models proposed in the literature \citep{melis2019resistance}.

A separate experiment was also conducted here in the same channel to determine $C_d$ for the steady and uniform flow case and for $m=194$. In these experiments, the staggered cylinder configuration was the same. The three target $S_o>0$ values were also used. The $C_d$ for these experiments determined from the load cells $C_{s,p}$ and separately from equation \ref{Manning_Sf} when setting $S_f=S_o$ are shown in Table \ref{Tab:Table1} for completeness. In both cases, $C_d$ was computed using the constricted cross-section velocity $U_c=U/(1-\sqrt{2\phi_v/\pi})$ \citep{etminan2017new}. 

\begin{figure}[ht!]
\noindent\includegraphics[width=20pc]{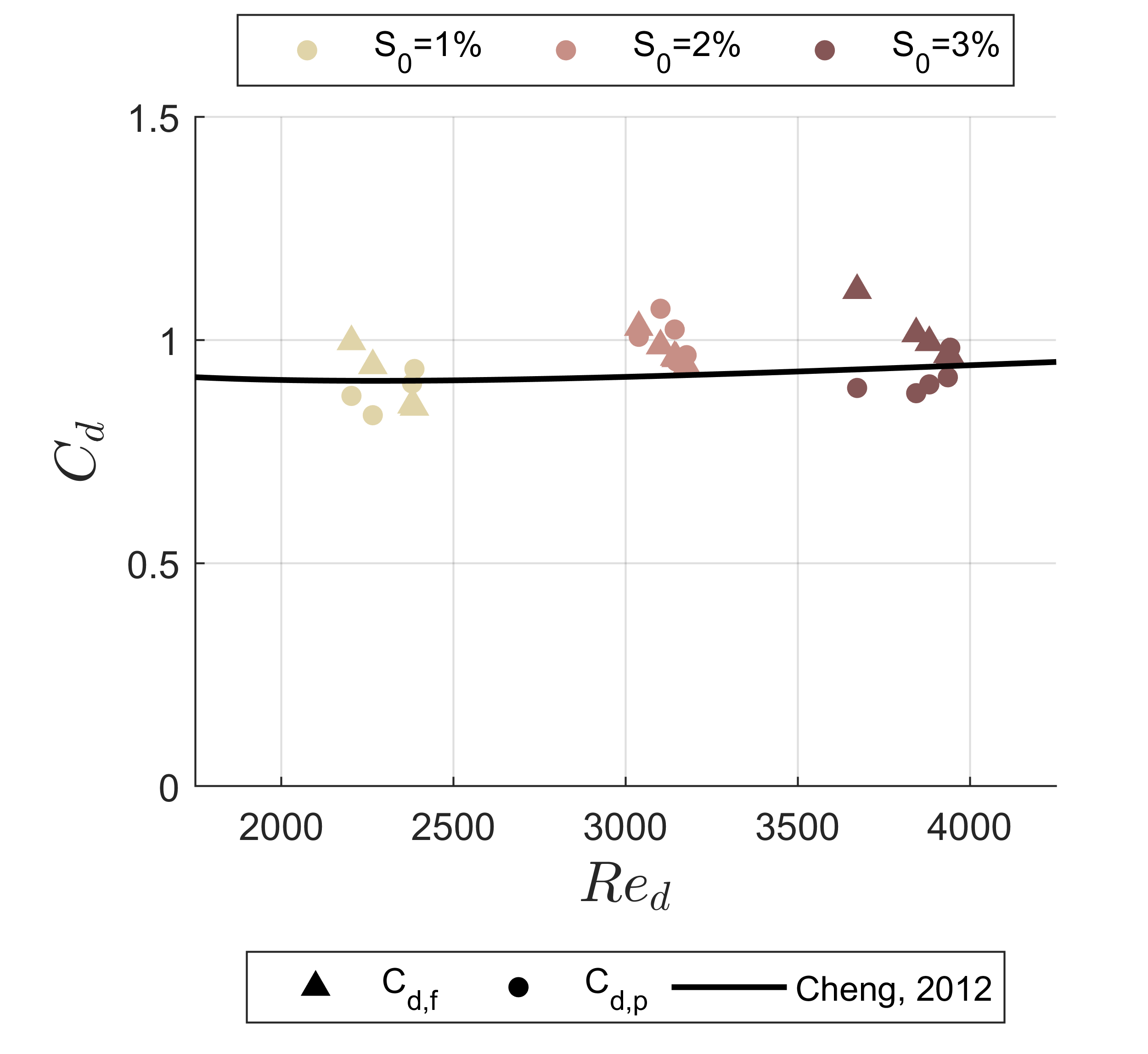}
\caption{The comparison between the drag coefficient $C_d$ derived for an isolated cylinder \citep{cheng2012calculation}, measured by the load cells $C_{d,p}$ and derived from friction slope $C_{d,f}$ using equation \ref{Force_Balance} - all presented as a function of Reynolds number $Re_d$ for $S_f=S_o>0$ when steady and uniform flow conditions are  attained.  The agreement between $C_{d,p}$ and $C_{d,f}$ support the assumption that wall friction can be ignored relative to the drag for such a rod density.}
\label{Fig:Cd uniform flow}
\end{figure}

\begin{table}[!ht]
    \caption{Summary of the drag coefficient measurements from the load cells and from the friction slope $S_f$ for steady-uniform flow (i.e. $S_f=S_o$) for different $S_o$.  Here, the $h$ is the mean water depth, $Q$ is the discharge, $U$ is the constricted cross-section velocity \cite{etminan2017new}, $F_d$ is the mean drag measured by the load cells, $C_{d,p}$ is the drag coefficient determined from the load cells, $C_{d,f}$ is the drag coefficient determined from the uniform flow result with $S_f=S_o$ (i.e. equation \ref{Force_Balance}), and $Re_d$ is, as before, the Reynolds number.  The overall agreement between the $C_d$ estimates by the two methods is better than $10\%$ on average.}    
    \label{Tab:Table1}
    \begin{center}
    \begin{tabular}{|c|c|c|c|c|c|c|c|c|c|}
    \hline
        $\  S_0\  [\%]\  $ & $\  h\  [m]\  $ & $Q\  [m^3/s]\  $ & $\  U\  [m/s]\  $ & $\  F_d\  [N]\  $ & $\  C_{d,f}\  $ & $\  C_{d,p}\  $ & $\  Re_d\  $  \\ \hline
        1 & 0.08 & 0.016 & 0.42 & 0.037 & 1.00 & 0.87 & 2204  \\
        1 & 0.10 & 0.021 & 0.43 & 0.046 & 0.94 & 0.83 & 2266  \\
        1 & 0.12 & 0.026 & 0.45 & 0.069 & 0.85 & 0.93 & 2387  \\
        1 & 0.14 & 0.030 & 0.45 & 0.077 & 0.85 & 0.90 & 2380  \\ \hline
        2 & 0.06 & 0.017 & 0.60 & 0.061 & 0.96 & 0.95 & 3147  \\
        2 & 0.08 & 0.022 & 0.58 & 0.080 & 1.03 & 1.01 & 3038  \\
        2 & 0.10 & 0.029 & 0.60 & 0.105 & 0.94 & 0.97 & 3177  \\
        2 & 0.12 & 0.034 & 0.60 & 0.131 & 0.96 & 1.02 & 3142  \\
        2 & 0.14 & 0.040 & 0.59 & 0.156 & 0.99 & 1.07 & 3101  \\ \hline
        3 &	0.06 & 0.020 & 0.70 & 0.078 & 1.11 & 0.89 & 3672 \\ 
        3 &	0.08 & 0.028 & 0.73 & 0.113 & 1.01 & 0.88 & 3844 \\
        3 &	0.10 & 0.035 & 0.74 & 0.147 & 0.99 & 0.90 & 3882 \\
        3 &	0.12 & 0.043 & 0.75 & 0.184 & 0.97 & 0.92 & 3935 \\
        3 &	0.14 & 0.050 & 0.75 & 0.231 & 0.96 & 0.89 & 3943 \\ \hline
    \end{tabular}
    \end{center}
\end{table}
\begin{figure*}[ht!]
\noindent\includegraphics[width=40pc]{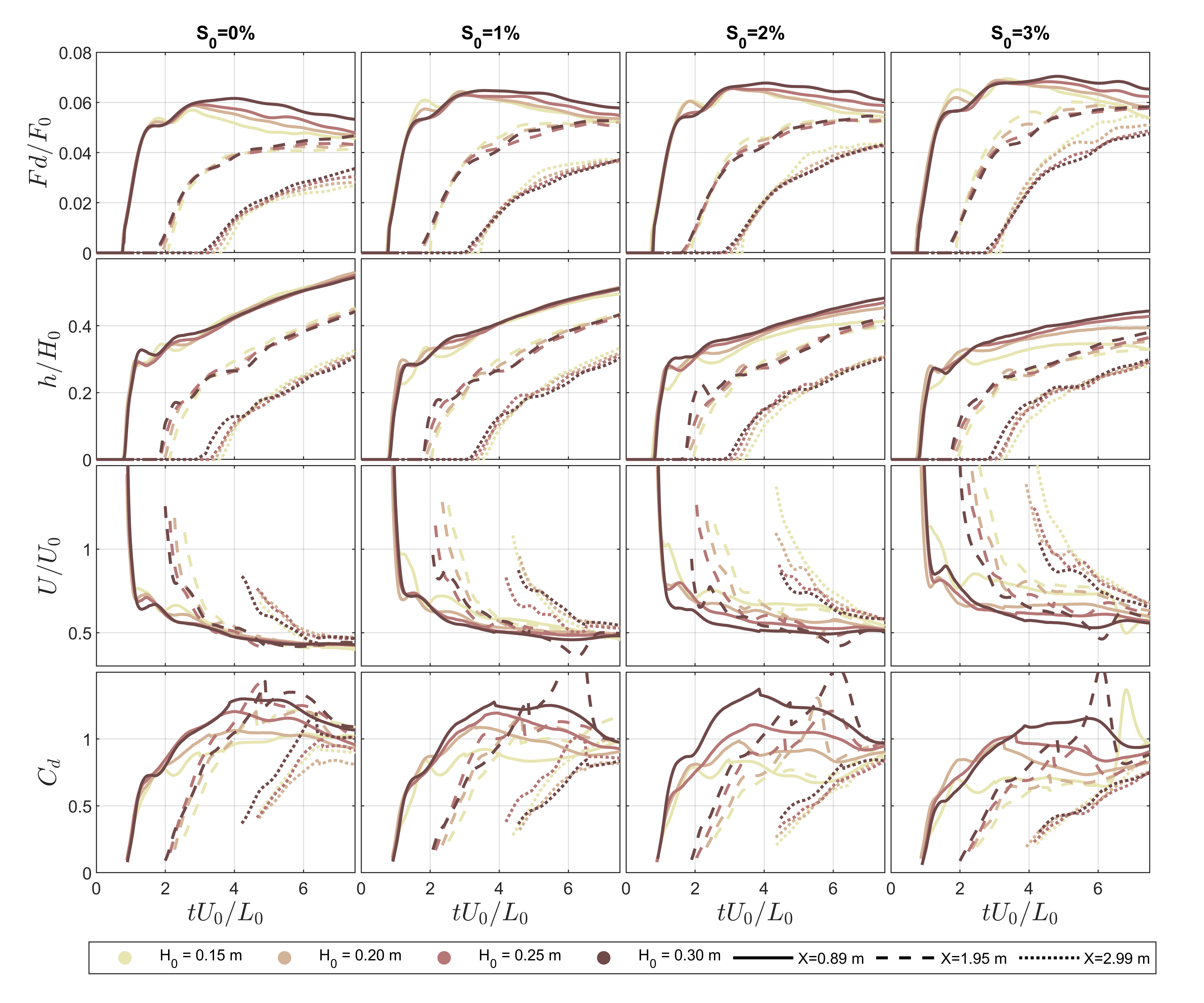}
\caption{Variations of the normalized drag, water level, velocity, and $C_d$ as a function of time at the three load cell locations for all $H_o$ and $S_o$ conditions (columns). Here, $F_o$ is the normalizing force computed using $C_d=1$, $U_o^2=g H_o$, and $\rho=1000$ kg m$^{-3}$.}
\label{Fig:Cd load plates}
\end{figure*}

\begin{figure*}[ht!]
\noindent\includegraphics[width=40pc]{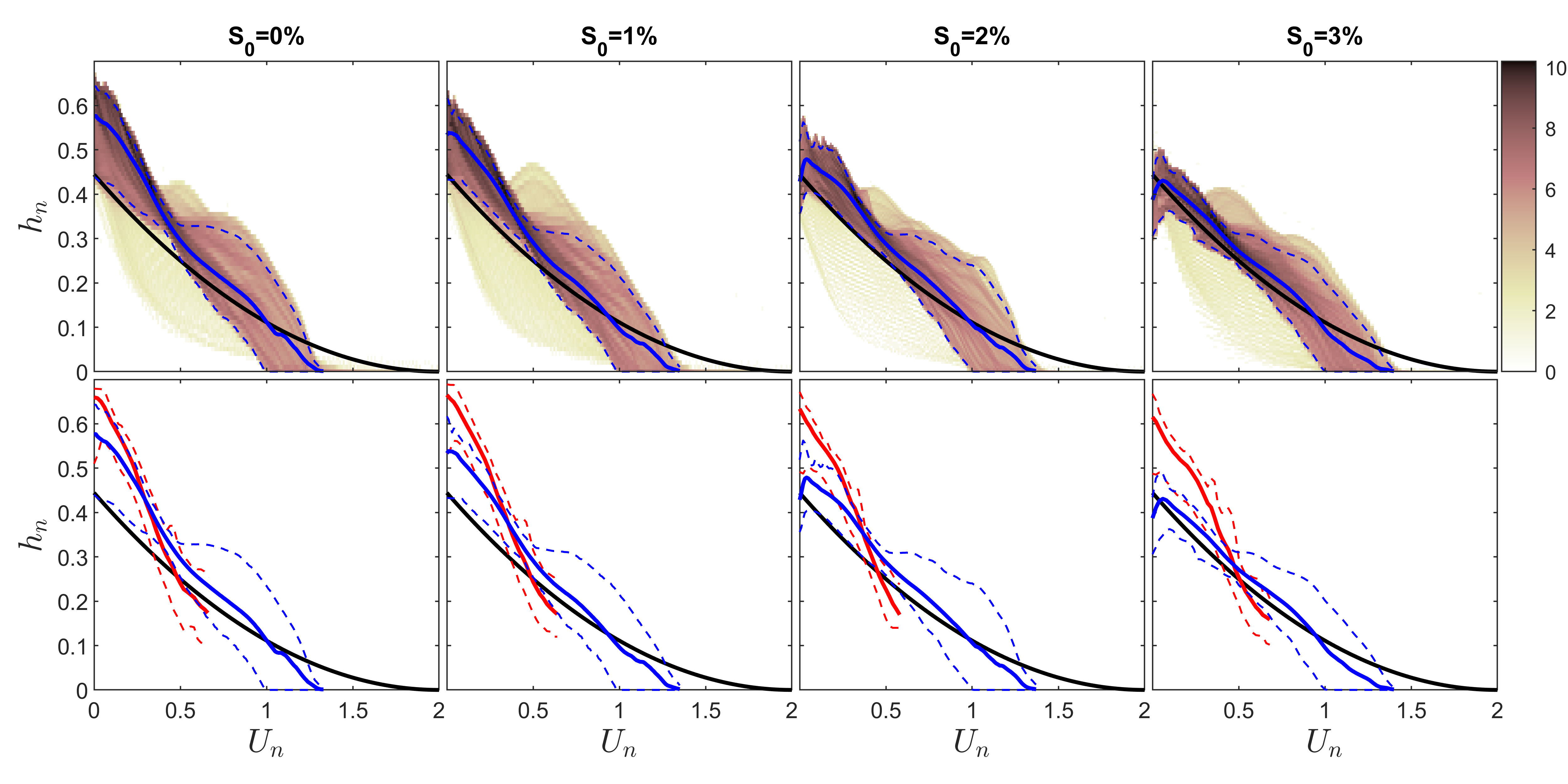}
\caption{The measured normalized water depth $h_n=h/H_o$ (ordinate) variations with the normalized velocity $U_n=(x/t)/U_o$ (abscissa) along with the Ritter solution (black solid line) for all four bed slopes $S_o$. In the top panels, the blue lines represent the derived $h_n$ (solid line) and the 95\% confidence bound (dashed lines). The color scale is featured to show the $log_{10}(N)$ data points used in the derivation of $h_n-U_n$ relation. The bottom panels repeat the blue lines ($m=194$) from above and add earlier experiments ($m=1206$) for the dense rod density case \cite{melis2019resistance} (red lines).}
\label{Fig:Ritter comparrison}
\end{figure*}

\section{Results and Discussion}
To address the study objectives, this section is organized as follows.  A comparison between measured $C_d$ for steady-uniform flow cases and dam-break cases for all $S_o$ values is presented.  The magnitude and controls on $C_d$ for this rod configuration for steady-uniform flow can be used to ascertain whether wall friction can be ignored relative to the canopy drag and whether $C_d$ estimates from the load cells match the expectations from equation \ref{Drag_Sf}.  Next, the effects of $H_o$, $S_o$ and $S_f$ on the wavefront are considered.  These considerations are also used to summarize the data from the experiments in a normalized manner.  Once again, data from the prior study ($m=1206$) and the present experiments ($m=194$) are compared to assess the effects of $C_d m D$ on $S_f$.  To facilitate comparisons across the $H_o-S_o$ cases and the two $m$ values, a single reference curve was repeated in all of them based on equation \ref{Ritter_n}.  This 'base-line' curve makes a logical choice for a reference because it is derived for $S_f=0$ and $S_o=0$. The physics of the advancing front wave is considered next.  Two regimes are shown to emerge when analyzing the measured front position $x_f$ against estimates from $U_o t$.  The first is a rapid regime dominated by both inertial and frictional effects, and a second regime trending towards the diffusive wave approximation where the frictional effects experience a reduced $C_d$.  A discussion as to the possible causes of this drag reduction is then offered. Throughout, the results in the figures are presented in dimensionless form using the following: water depths are normalized by $H_o$, velocities are normalized by $U_o$, time is normalized by $U_o/L_o$ where $L_o=(C_d m D)^{-1}$ is a reference adjustment length \citep{belcher2003adjustment} taken at $C_d=1$ (i.e. steady-uniform flow case), forces are normalized by $(1/2) C_d \rho D H_o U_o^2$ and horizontal distances are normalized by $L_o$.   

\subsection{Drag Coefficient from the Load Cells}
For steady-uniform flow, the measured $C_d$ from the load cells $C_{d,p}$ and from friction slope $C_{d,f}$ shown in Figure \ref{Fig:Cd uniform flow} do not deviate significantly from the accepted formulation for an isolated cylinder $C_{d,iso}$ that is given by \citep{cheng2012calculation,wang2015steady}
\begin{equation}
\label{Cd_iso}
C_{d,iso}=11 (Re_{d})^{-0.75}+0.9{{\Gamma }_{1}}\left( {{{Re}}_{d}} \right)+1.2{{\Gamma }_{2}}\left( {{{Re}}_{d}} \right),
\end{equation} 
where $Re_d=U_c D/\nu$ is the element Reynolds number and  
\begin{align}
\label{gamma1}
{{\Gamma }_{1}}\left( {{{Re}}_{d}} \right)=&1-\exp \left( -\frac{1000}{{{{Re}}_{d}}} \right), \nonumber \\
{{\Gamma }_{2}}\left( {{{Re}}_{d}} \right)=&1-\exp \left[ -{{\left( \frac{{{{Re}}_{d}}}{4500} \right)}^{0.7}} \right].
\end{align} 
This expression assumes that $Re_d<10^4$ (and is below the drag crisis range for isolated cylinders) and that the drag from each cylinder operates in isolation (i.e. no interference, sheltering, or blocking).  The agreement between $C_{d,p}$ and $C_{d,f}$ is also suggestive that wall friction that impacts $C_{d,f}$ but not $C_{d,p}$ can be ignored relative to the canopy drag for such a rod density.  In the range of $Re_d$ considered here, a $C_d=0.9-1.1$ appears to describe the steady-uniform flow data without any significant dependency on $Re_d$ as shown in Figure \ref{Fig:Cd uniform flow}.  For this reason, the reference drag $C_d=1$ is selected in the $L_c$ calculations used for normalizing longitudinal distances.  Returning to the dam break cases, the load cell measured $F_d$ was used, together with the imaged water depth $h$, to compute $C_d$ using $F_d= \left(1/2\right) C_d \rho D h U^2$.  Here, the $U$ was computed from the continuity equation using the imaged water depth as shown in Figure \ref{Fig:Cd load plates}. The computed $C_d$ is well below unity as the wave front passes.  This reduction in $C_d$ is significant for all $S_o$ values. For early times $t (U_o/L_o)<1$, $C_d$ increases from some 0.2 to 0.6 as shown in Figure \ref{Fig:Cd load plates} with a mean of about $C_d=0.4$. The mean value here is consistent with the value inferred indirectly from fitting the SVE to measured $h(x,t)$ in the prior study \citep{melis2019resistance} despite the large difference in $m$ between the two experiments. The agreement between the reduced drag value ($C_d=0.4$) across the two experiments hints that sheltering alone may not be the main mechanism responsible for drag reduction as sheltering is expected to dependent on rod density.
\begin{figure*}[ht!]
\noindent\includegraphics[width=40pc]{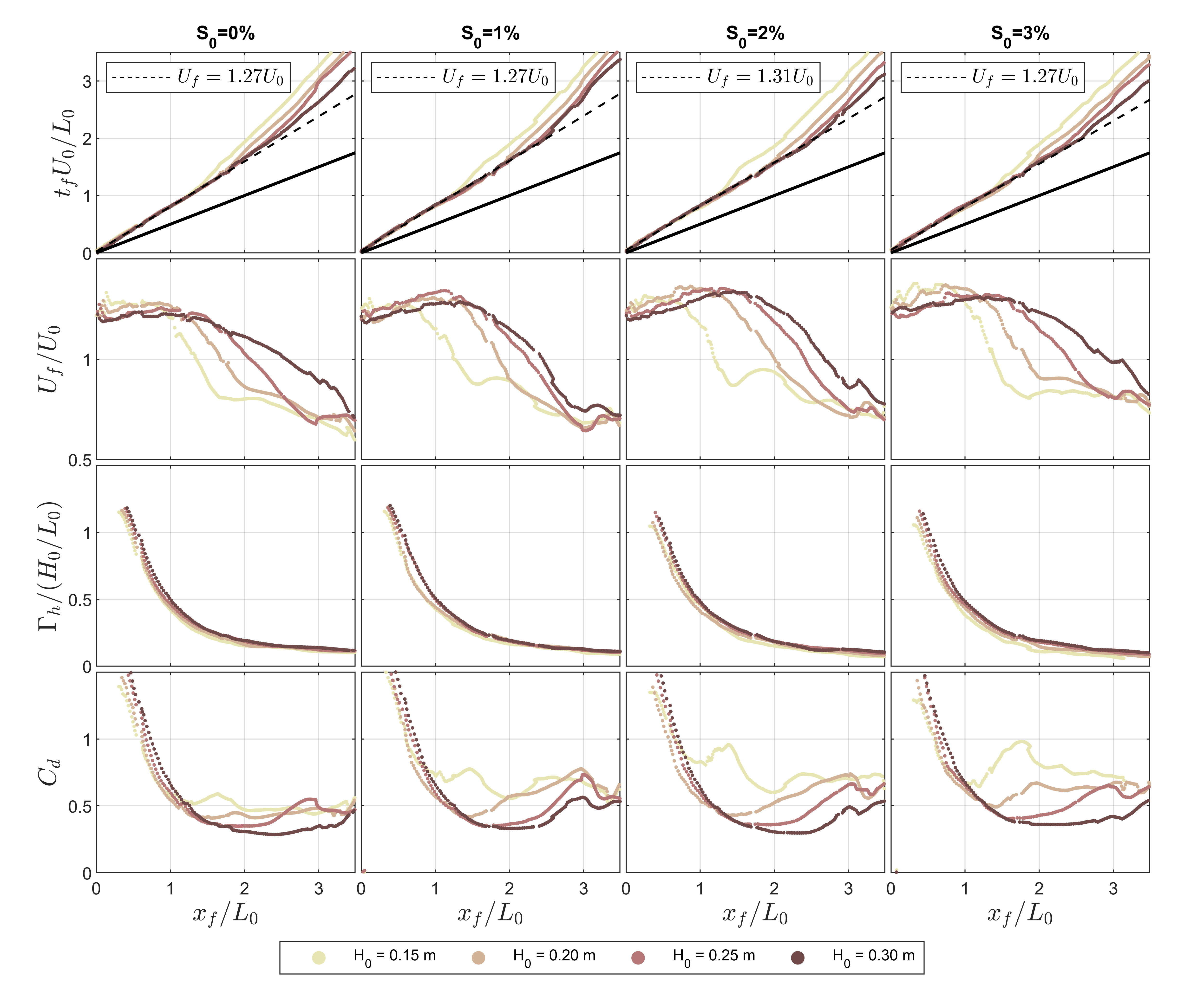}
\caption{The relation between wave front position $x_f$ and those determined when assuming $x_f=U_o t$ from the top to the bottom: the time $t_f$ multiplied by the characteristic velocity $U_o$, the normalized wave front velocity $U_f/U_o$, the tip slope $\Gamma_h$ divided by characteristic depth $H_o$ and the inferred drag coefficient $C_d$ using the diffusive wave approximation.}
\label{Fig:Advancing tip}
\end{figure*}

\subsection{Bed Slope and Frictional Effects}
To illustrate the simultaneous effects of $S_o$ and $H_o$ variations on the depth-velocity relations, the experiments here ($m=194$) are summarized in Figure \ref{Fig:Ritter comparrison} and then compared to equation \ref{Ritter_n} for $S_o=S_f=0$.  The prior experiments for $m=1206$ are also added for reference and are organized, as before, by the two control variables $S_o$ and $H_o$.  A number of comments can be made about Figure \ref{Fig:Ritter comparrison}:

\begin{itemize}
\item Equation \ref{Ritter_n} over-predicts the advancing front wave speed (i.e. the $U_n$ associated with $h_n<0.05$) for all $S_o-H_o$ cases compared to their frictional counterparts as expected; 
\item The higher $m$ experiments result in higher water pile-up (i.e. higher $h_n$) at the smaller $U_n<0.5$ values as expected; 
\item With increasing $S_o$, the lower $m$ data approaches equation \ref{Ritter_n} (i.e. the solution for $S_o=S_f=0$) except for the advancing wave front region.  This finding may be explained by the fact that increasing $S_o$ also increases $S_f$ thereby diminishing their difference in the SVE ($S_f$ is always finite and large in the presence of a canopy).  The advancing front (i.e. the region with $h_n<0.05$) always experiences a slow-down (lower $U_n$) compared to equation \ref{Ritter_n} as noted earlier.  
\end{itemize}
In sum, Figure \ref{Fig:Ritter comparrison} confirms all the logical expectations of the $u_n-h_n$ relations derived from the experiments for differing $m$ values, $S_o$, and $H_o$. Common to all the cases is the slow-down for the advancing front region when compared to the $S_f=0$ case due the canopy.  In the absence of a canopy, the Ritter solution reasonably describes the advancing wavefront for the channel setup here as discussed elsewhere \citep{melis2019resistance}. The advancing front region is now explored in detail.

\subsection{Two Dynamical Regimes}
The dynamical regimes that introduce deviations from equation \ref{Ritter_n} are examined along $x$.  These regimes are identified by exploring how the normalized front position $x_f/L_o$ varies against $t U_o/L_o$ as shown in Figure \ref{Fig:Advancing tip}. Two distinct regimes can be identified with a transition governed by the initial $H_o$ for all $S_o$ (top panels).  At small distances from the dam (first regime, $x_f/L_o<$ 1), a robust $x_f=c_f U_o t$ can be seen from the experiments where $c_f=1.26$ for $S_o=0$ and $c_f=1.33$ for $S_f=3\%$.  This implies that the advancing front velocity $U_f = c_f \sqrt{g H_o}$ is roughly a constant and weakly dependent on $S_o$ given the small increase in $c_f$ with increasing $S_o$.  Setting $x/t=U$ in equation \ref{Ritter_n} leads to a frictionless advancing front speed that is almost twice as fast (i.e. $c_f=2$ for $S_f=S_o=0$) as discussed elsewhere \citep{hogg2004effects}. Beyond a near constant $U_f$ for $x_f/L_o<1$, another dynamically interesting feature of this regime is the rapid drop in measured $\partial h/\partial x$ with increasing $x_f$ until the attainment of the second regime $x_f/L_o>1$.  The second regime marks a gradual slow-down in $U_f/U_o$ compared to the near-constant velocity in the first regime (top two rows in Figure  \ref{Fig:Advancing tip}), but trends towards a near constant $\Gamma_h=-\partial h/\partial x$ with increasing $x_f/L_o$. Thus, while the first regime experiences a near constant $U_f$ and a variable $\Gamma_h$, the second regime is dynamically the opposite.  The increase in $H_o$ delays the onset of the slow down (or second) region with increased $x_f$ (i.e. reduced $U_f$) for all $S_o$ values.   It is expected that with further increases in $x_f/L_o$, $U_f/U_o$,  $\Gamma_h L_o/H_o$, and likely $C_d$ all attain a constant value in agreement with the diffusive wave approximation.  The measurements reported in Figure \ref{Fig:Advancing tip} seem to support this extrapolation for the low $H_o$ cases considered.

The fact that $U_f$ varies may appear counter to the approximated physics in equation \ref{SVE_Diffusive_Wave}.  However, a detailed analysis (not shown here) conducted on the data suggest that 
\begin{equation}
\label{approximated_SVE}
\frac{\partial U h}{\partial x}\approx U_f \frac{\partial h}{\partial x}; \left | \frac{\partial U}{\partial t}+U\frac{\partial U}{\partial x}\right |<< \left| \frac{\partial h}{\partial x}\right|. 
\end{equation} 
0Thus, the diffusive wave approximation remains plausible in the second regime despite variations in $U_f$ with $x_f$.  

Returning to the drag reduction issue, a $C_d$ was inferred from equation \ref{SVE_Diffusive_Wave_Speed} using measured $U_f$ and measured $\partial h/\partial x$ for the $S_o-H_o$ combinations.  These computed $C_d$ values are shown in Figure \ref{Fig:Advancing tip}.  To be clear, the estimate of $C_d$ using a diffusive wave approximation for the first region (i.e. $x_f/L_o<1$) cannot be correct.  The computed $C_d$ is only presented here to corroborate the transition zone from the first to the second region around $x_f/L_o=1$, where the diffusive wave approximation begins to apply for $x_f/L_o>1$.  Beyond that $x_f$, a near constant $C_d$ is attained from equation \ref{SVE_Diffusive_Wave_Speed}. The low $C_d$ values measured by the load cells shown in Figure \ref{Fig:Cd load plates} agree with those computed from the diffusive wave approximation using measured $h(x,t)$.  This agreement in reduced $C_d$ for the second region serves two purposes: (i) it shows that the diffusive wave approximation is plausible for $x_f/L_o>1$, and (ii) it confirms that a new drag reduction mechanism must be operating that is unique to the dam break problem (i.e. it does not exist in the steady-uniform flow cases).

\subsection{What Causes the Drag Reduction?}

\begin{figure}[ht!]
\noindent\includegraphics[width=20pc]{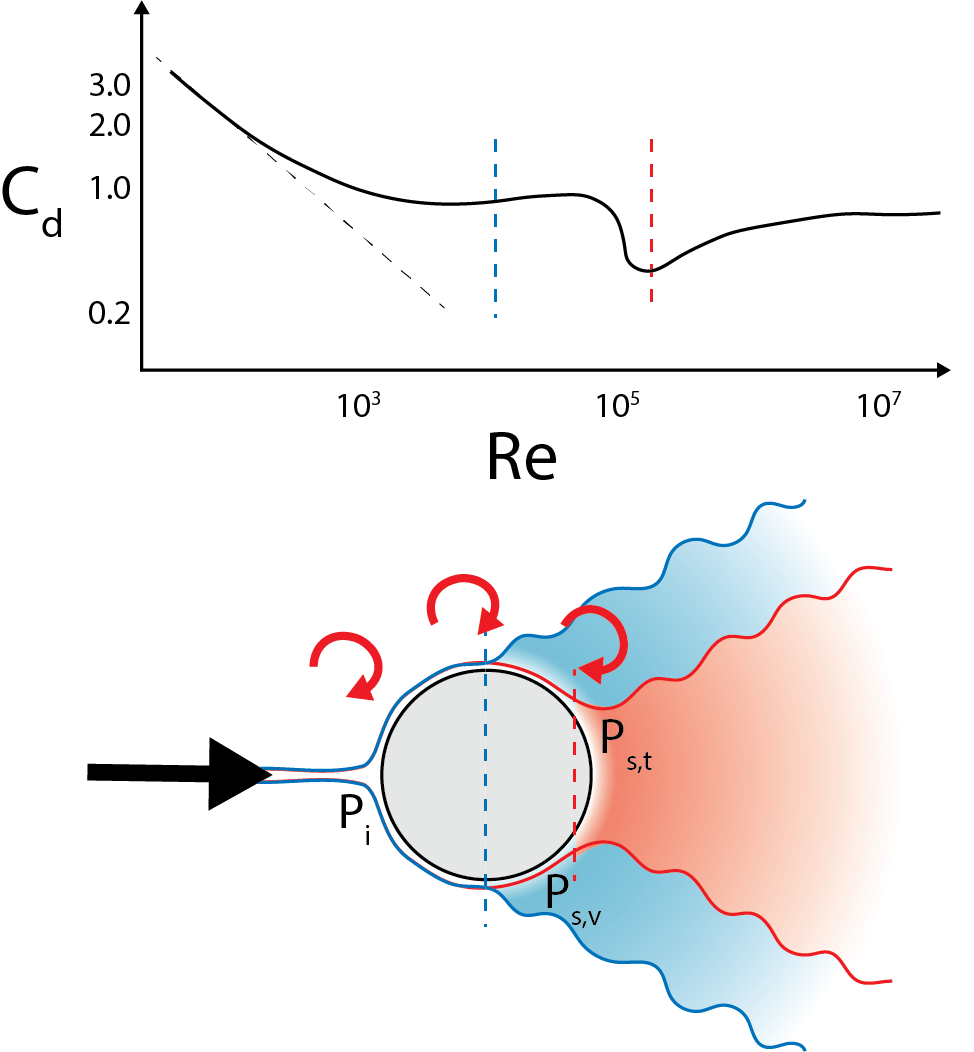}
\caption{Conceptual model for the boundary layer detachment from a cylinder during the drag crisis (red wake, boundary layer is turbulent) compared with pre-crisis (blue wake, boundary layer is laminar).  The delayed separation during the drag crisis has two effects: (i) it allows for a pressure recovery (i.e. $P_{s,t}$ is less negative than $P_{s,v}$ shown at the dashed blue center line of the cylinder), thereby reducing the pressure difference between the front ($P_i$) and back of the cylinder (i.e. $P_{s,v}$ or $P_{s,t}$) and (ii) it reduces the wake area behind the cylinder (red vs blue).  Both effects act in concert to reduce the form drag $C_d$ shown on top as a function of $Re$.  The dashed vertical lines show the $Re$ for the pre-crisis (blue) and during the drag crisis (red).}
\label{Fig:DragCrisis}
\end{figure}

To recap the findings thus far, the $m=1206$ experiments yielded a $C_d=0.4$ inferred from $h(x,t)$ measurements by fitting an optimum $C_d$ to the SVE so as to match $h(x,t)$. It was postulated that this fitted and reduced drag is linked to the so-called drag crises \citep{melis2019resistance}.  For an isolated cylinder immersed in a steady and uniform background flow that describes its far-field, the drag crisis occurs when well-organized vortex shedding (e.g. Karman-vortex streets) are disrupted and transition to randomized shedding with further increases in $Re_d$ \citep{singh2005flow}. This transition occurs at very large $Re_d (>10^5)$ in the isolated cylinder case and for steady-uniform flow as shown in Figure \ref{Fig:DragCrisis}.  Numerical simulations confirm that the drag crisis commences when a critical Reynolds number is reached where the boundary layer on the cylinders become turbulent thereby maintaining attachment to the cylinder further downstream \citep{singh2005flow}.  The transition to a turbulent boundary layer and a sustained attachment onto the cylinder has two effects.  The first is that the pressure differential between the front ($P_i$) and the back ($P_{s,t}$ for turbulent versus $P_{s,v}$ for laminar) of the cylinder is reduced due to the partial pressure ($P_{s,t}>P_{s,v}$) recovery following a longer downstream attachment \citep{wen2017investigation} along the back of the cylinder as schematized in Figure \ref{Fig:DragCrisis}.  Numerical simulations for an isolated cylinder with a uniform far field flow have shown that this pressure recovery does occur because of delayed separation on the back of the cylinder and can reduce the pressure differential (referenced to $P_i$) by more than $25\%$. The second is that the wake area is also reduced (red versus blue shades in Figure \ref{Fig:DragCrisis}) as detailed by simulations \citep{singh2005flow,wen2017investigation}.  The combined effect of reduced pressure differential and reduced wake area leads to a drastic reduction in $C_d$ (more than a factor of 2) - or the drag crisis.

It was conjectured in the prior study \citep{melis2019resistance} that the dam break problem leads to a disturbed and transient 'far-field' background flow region that enables the randomization of vortex shedding to be initiated and persistent at much lower $Re_d$.  This disturbed far-field state can enhance momentum transport to the turbulent boundary layer region attached onto the cylinder and allows the drag crisis to be maintained as schematized in Figure \ref{Fig:DragCrisis}). A reduced $C_d$ is then to be sustained over an extended range of $Re_d$ well below the critical value of $Re_d=10^5$ (where the laminar boundary layer flips to a turbulent state for an undisturbed background state).   

\begin{figure*}[ht!]
\noindent\includegraphics[width=40pc]{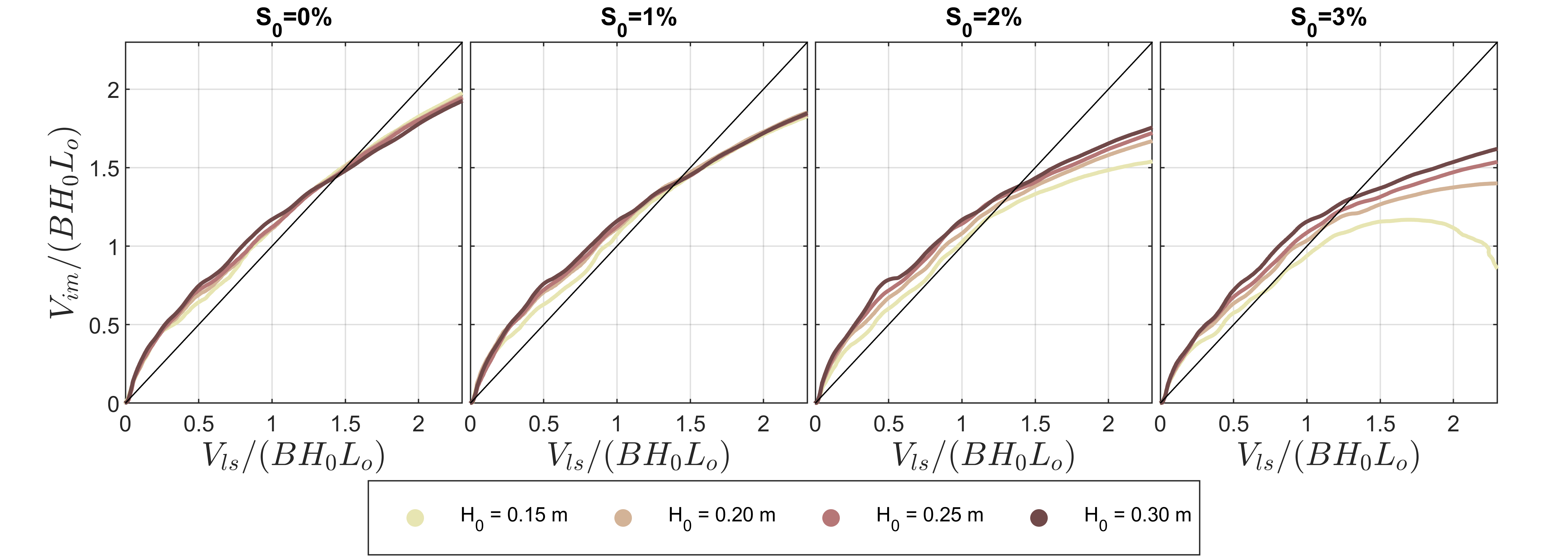}
\caption{A comparison between normalized water volume within the channel for $t>$ 0 from the imaging system (ordinate) for $x>0$ and from level sensors (abscissa) behind the dam $x<0$.}
\label{Fig:Air entrance}  
\end{figure*}

The work here offers an amendment to this argument, which is the air entrainment leading to density reductions at the advancing front in the second region as shown in the photographs of Figure \ref{Fig:setup}.  Figure \ref{Fig:Air entrance} presents a comparison between the measured normalized volume of the inflow behind the dam using water level sensors (ordinate) and the imaged volume of water as the front progresses downstream from the dam for $x/t>0$ (abscissa).  Noting that $x_f \approx 1.3 U_o t$ in Figure \ref{Fig:Advancing tip} and that the test-section region analyzed in Figure \ref{Fig:Advancing tip} is for $x_f/L_o \leq 3$, it is clear that the imaged outflow volume appears to be consistently larger by some 10-20$\%$ compared to the inflow volume (i.e. volume behind the dam measured by detailed water level measurements).  From an experimental uncertainty point of view, the splashing and breakup of water into fine droplets near the advancing tip region shown in Figure \ref{Fig:setup} would have reduced, not increased, the imaged outflow volume. Water droplets, which are not counted in the imaged $h(x,t)$ volume calculations, will be missed.  Since water mass is conserved, the imaged outflow volume exceeding the inflow volume must then be associated with some air entrained at the advancing front.  With air volume entrained at the advancing wave front region, the overall water density near the tip front must be substantially reduced.  A reduction in water density near the advancing front region leads to a concomitant overall inferred $F_d$ and $C_d$ reductions in this vicinity.  This drag reduction mechanism may be acting in concert with the randomization of vortex shedding associated with the earlier speculated drag crisis \citep{melis2019resistance}.  Thus, air entrainment, reduced water density at the advancing front region, and a disturbed background state all conspire to reduce $C_d$ by a factor of $2$.

\section{Conclusions and Broader implications}
\label{Conclusions}
The closure of the friction slope in the SVE continues to draw research attention as it encodes all the solid-fluid interactions.  In operational models of flood waves, the $S_f$ is related to $U^2/(2 g)$ using conventional formulations that aim to recover the steady-uniform flow (e.g. Manning's formula).  However, the presence of drag elements adds another layer of complexity and that cannot be represented as an equivalent wall friction derived using steady-uniform flow conditions.  Those interactions are represented by a drag coefficient $C_d$ that modifies $S_f$ as shown here.  As before, steady-uniform flow are assumed as base line state to link $C_d$ to $S_f$ in practice. This approach was deemed pragmatic in many dam break and flood routing applications because there are numerous data sets and models for $C_d$ when the drag elements are rod canopies (or other approximated geometries).  Much research has focused on corrections to $C_d$ from an isolated cylinder and include sheltering, blockage, among others.  The work here suggests that $C_d$ from steady-uniform flow may be an overestimate by a factor of 2.  New physics at the advancing front related to air entrainment and randomization of coherent vortex structures occur due to the unsteady and disturbed nature of the flow away from the rod canopy.  These two effects act as drag reduction mechanisms that 'speed up' the advancing front relative to $C_d$ models derived from steady-uniform flow.  

The work also identified two dynamically interesting regimes for the advancing front velocity based on where the front location $x_f$ is relative to the adjustment length scale $L_o=(C_d m D)^{-1}$. For $x_f/L_o<1$, the front velocity is roughly constant and scales with $\sqrt{g H_o}$ and $\partial h/\partial x$ rapidly declines in magnitude.  A second regime for $x_f/L_o>1$ emerges where the front velocity begins to decline but  $-\partial h/\partial x$ begins to attain its minimum.  The diffusive wave approximation further suggests that in the second regime, the drag coefficient is reduced by a factor of 2 relative to its steady-uniform value. 

This work must be viewed as an embryonic step so as to improve flood forecasting in the future. Additionally, the present results offer benchmark data for future numerical investigations given the increased focus on modeling and simulating dam-break problems \cite{akkerman2012free,dai2022numerical,wu1999variation,ai2022three,hernandez2023bathymetry,eames2022horizontal,yan2023fast} over obstructions and vegetated bed scenarios \cite{he2017numerical,khoshkonesh2022dam,zhang2016numerical,khoshkonesh2023study}. The need to move beyond wall friction representation for energy losses is recognized in large scale models but alternatives remain in short supply \citep{melis2019resistance}. Progress on these alternatives using a quadratic drag law is timely given the rapid advancements in water level measurements from space (to within 10 cm)  \citep{alsdorf2007measuring} and the wealth of research on drag coefficients for different geometries as derived from steady-uniform flow.   However, the disturbed and transient nature of this flow was shown to lead to previously unexplored drag reduction mechanisms (air entrainment, drag crisis). Future experiments should seek novel methods to characterize the water density reductions and the randomization of vortex shedding associated with the drag crisis for such disturbed flows.  From the simulation perspective, the results here hint that a 3-phase representation (solid, water, and air) may be needed to capture the interplay between air entrainment and boundary layer separation at the solid interface of the cylinders during the dam break. 

\begin{acknowledgments}
EB acknowledges Politechnico di Torino (Italy) for supporting the visit to Duke University. GK acknowledges support from the U.S. National Science Foundation (NSF-AGS-2028633) and the U.S. Department of Energy (DE-SC0022072).   DP acknowledges support from Fondo europeo di sviluppo regionale (FESR) for project Bacini Ecologicamente sostenibili e sicuri, concepiti per l'adattamento ai Cambiamenti ClimAtici (BECCA) in the context of Alpi Latine COoperazione TRAnsfrontaliera (ALCOTRA) and project Nord Ovest Digitale e Sostenibile - Digital innovation toward sustainable mountain (Nodes - 4). 
\end{acknowledgments}

\bibliography{main_bib_DamBreak2024}

\section*{Conflict of Interest Statement}
The Authors have no conflicts to disclose.

\section*{Data availability statement}
The data that support the findings of this study are available from the corresponding author upon reasonable request.

\end{document}